\documentclass[%
 aip,
 amsmath,amssymb,
reprint,%
]{revtex4-1}

\usepackage{graphicx}
\usepackage{dcolumn}
\usepackage{bm}

\usepackage[utf8]{inputenc}
\usepackage[T1]{fontenc}
\usepackage{mathptmx}
\usepackage{color,soul}

\usepackage{xcolor}
\definecolor{knof1}{HTML}{193E73}
\definecolor{knof2}{HTML}{ff5B00}
\definecolor{knofgreen}{rgb}{0.56,0.69,0.19}
\definecolor{knofpurple}{rgb}{0.57,0.0,0.15}
\begin{document}

\preprint{AIP/123-QED}

\title[Aperiodic electro-optic time lens for spectral manipulation of single-photon pulses]{Aperiodic electro-optic time lens for spectral manipulation of single-photon pulses}

\author{F. So\'snicki}
 \email{Filip.Sosnicki@fuw.edu.pl}
\author{M.  Miko\l{}ajczyk}%
\author{A. Golestani}
\author{M. Karpi\'nski}
\affiliation{ 
Faculty of Physics, University of Warsaw, Pasteura 5, 02-093 Warszawa, Poland
}%

\date{\today}

\begin{abstract}
Electro-optic time lenses are promising experimental components for photonic spectral-temporal processing of quantum information. We report a stable method to realize an electro-optic time lens, which relies on the amplification of an electronic response of a fast photodiode. The method does not require a repetitive clock and may be applied to aperiodic optical signals. We experimentally demonstrate the approach using single-photon pulses, and directly verify its aperiodicity. The approach will enable construction of complex electro-optic temporal optical systems.
\end{abstract}

\maketitle

Recent years have shown an extensively increasing interest in controlling quantum light in the spectral-temporal degree of freedom. It has attracted the attention of the quantum information processing community as it may pave the way towards multi-dimensional quantum information processing on a large scale\cite{Wasilewski07,Roslund13,Brecht15,Kues17, Lukens17, Davis18, Raymer20}. It is naturally compatible with already existing telecom fiber infrastructure and with integrated photonic systems, where encoding quantum information in polarization or spatial modes is inefficient or challenging technically. To utilize the time-frequency domain for quantum information encoding, one needs to be able to shape both the temporal profiles as well as the spectra of single-photon pulses.

Electro-optic methods have been recognized as a promising approach to spectral-temporal manipulation of quantum light\cite{Kolchin08,Olislager12,Zhu13,Fan16,Karpinski17,Mittal17,Kues17,Sosnicki18,Lu18splitter}. Electro-optic phase modulation requires an RF driving signal which maintains timing correlation with the optical signal. In the case of quantum photonic applications, where the optical signal is in the form of single-photon pulses, exceptional timing stability is required. This is because eliminating loss and photon noise is a key concern in quantum photonic applications, which severely restricts the possibility of using feedback loops and multiplexed optical stabilizing signals.

The electro-optic time lens is realized by quadratic time-dependent phase modulation. It is a key building block for spectral-temporal manipulation of light \cite{Torres11, Salem13}. In particular, it forms the basis for spectral mode matching for quantum network applications,\cite{Karpinski17,Allgaier17,Sosnicki18} allows for spectral-temporal entanglement manipulation \cite{Mittal17} and may have applications in quantum key distribution \cite{Nunn13} and continuous variable quantum optics.\cite{Patera15} Among the possible approaches to time lens generation \cite{Torres11, Salem13, Lavoie13, Matsuda16, Allgaier17, Mazelanik20} the electro-optic approach stands out with its complete lack of optical noise and a high degree of wavelength tunability (no strong phase-matching constraints). In the standard approach to electro-optic time lens generation, the quadratic temporal phase modulation is approximated by a part of a high-bandwidth sinusoidal signal \cite{Torres11}. Generating the necessary stable sinusoidal signal requires phase-locking to a high-quality clock signal, which may not be readily available, in particular in a quantum photonic setting. Furthermore, the bandwidth of a phase-locking mechanism may restrict the ability of the setup to follow drifts in the optical path. Finally, sinusoidal approximations of time lenses are strictly periodic: they are only able to realize one time lens for every period of the sinusoid. This may be a limitation in quantum network applications, where single-photon signals of varying periodicities may need to be processed.

In this work, we present an experimentally simple, intrinsically stable approach for electro-optic time lens generation, which is well-suited for quantum photonic applications. Our approach relies on generating the quadratic time-dependent signal by directly amplifying the electronic output waveform of a fast photodiode. Apart from excellent timing stability, the approach allows aperiodic time lens generation, i.e.\ generating time lenses without a fixed repetition period. We experimentally demonstrate the approach by using the time lens to perform stable compression of the spectral bandwidth of telecom single-photon pulses over 24 hours without any feedback. Further, we experimentally verify the aperiodic property of the approach by spectrally compressing the bandwidth of pulses with a variable repetition period. We expect the approach to enable implementation of mode-matching devices for quantum network applications. Further, it will form a building block for multi-element temporal optical setups for more complex spectral-temporal manipulation of quantum light pulses.

Let us start by reviewing the concept of time lensing. It can be explained on~the basis of the optical space-time duality (OSTD) \cite{Kolner94,Torres11,Salem13}, which arises from the mathematical equivalence of equations describing paraxial diffraction of a beam confined in space and propagation of a narrowband optical pulse in dispersive medium within the slowly varying envelope approximation. There, chirping an~optical pulse in~a~dispersive medium, described by applying a~quadratic spectral phase $\Phi (\omega-\omega_{0})^{2}/2$, where $\Phi$ is~group delay dispersion (GDD), is a spectro-temporal analogue of broadening a beam in space under diffraction, which is described by applying a quadratic phase in~spatial frequencies. Using the~OSTD, one can then postulate a~time lens (TL), which applies a time-varying quadratic phase $Kt^{2}/2$, to~the~optical pulse, where $K$ is a constant chirping factor, and $t$ is the time in a reference frame traveling with the center of the pulse. It corresponds to~a~quadratic, transverse spatial phase acquired across a~spatial beam upon passing through a lens. The OSTD is the basis for multiple spectral-temporal shaping paradigms both in classical and quantum photonic settings. Spectral bandwidth compression is a time-lens-based transformation of relevance for quantum network applications \cite{Karpinski17}. It is an OSTD analogue of collimating an approximately Gaussian beam (i.e.,\ reducing its spatial frequency spectrum) by placing a spatial lens one focal length away from the beam waist. Thus spectral bandwidth compression can be realized by chirping an optical pulse in a GDD medium, followed by a time lens with chirping factor fulfilling the collimation condition $K=\Phi^{-1}$.

\begin{figure}[t!]
\includegraphics[width=\linewidth]{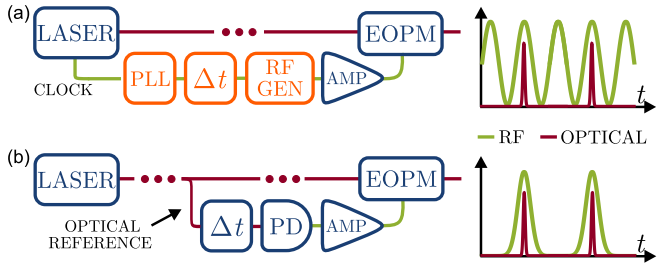}
\caption{\label{fig:ogolny} Conceptual schematic of periodic (a) and aperiodic (b) time lensing. The periodic time lens (a) is a standard method using an RF signal generator (RF GEN) synchronized to the laser by means of a phase-lock loop (PLL) using a clock signal from a laser. The generated signal is a sine, which is amplified (AMP) and drives an electro-optis phase modulator (EOPM), as shown in the panel on the right. The aperiodic time lens (b) is a much simpler method, where the signal used for modulation is derived directly from the laser pulses using a fast photodiode (PD), followed by an amplifier. $\Delta t$ -- synchronization.}
\end{figure}

To experimentally realize the time lens one needs to apply a parabolic radio-frequency (RF) signal to a traveling-wave electro-optic phase modulator (EOPM). The standard approach is to approximate the parabola with an extremum of a sinusoid. This is done by amplifying a single-tone RF signal, generated with an RF signal generator, and synchronizing the optical pulses passing through the EOPM with the minimum or maximum of the sinusoid, as shown in Fig.~\ref{fig:ogolny}(a). This results in applying a quadratic temporal phase with chirping factor $K = 4\pi^{2} f_{\mathrm{RF}}^{2}\,A$, where $f_{\mathrm{RF}}$ is the RF signal frequency and $A$ is its amplitude \cite{Torres11}. In this approach, a clock signal synchronized with optical pulses is required to drive a phase-lock loop (PLL) necessary to maintain a stable synchronization of the RF signal with the optical pulses. The modulation frequency $f_{\mathrm{RF}}$ has to be chosen as a harmonic of the repetition frequency of the optical pulses $f_{L}$, such that every optical pulse is modulated in the same way. Therefore this method is applicable only to periodic optical signals. In particular, it cannot be used to modulate double pulses, i.e.,\ two pulses within one repetition period $f_{\mathrm{L}}^{-1}$, where one of them is delayed with respect to the other by $\Delta \tau \neq n f_{\mathrm{RF}}^{-1}$, where $n$ is an integer. In such a case, the pulses encounter different parts of the sine signal, so only one pulse of the pair acquires the desired quadratic phase.

\begin{figure*}
\includegraphics[width=\textwidth]{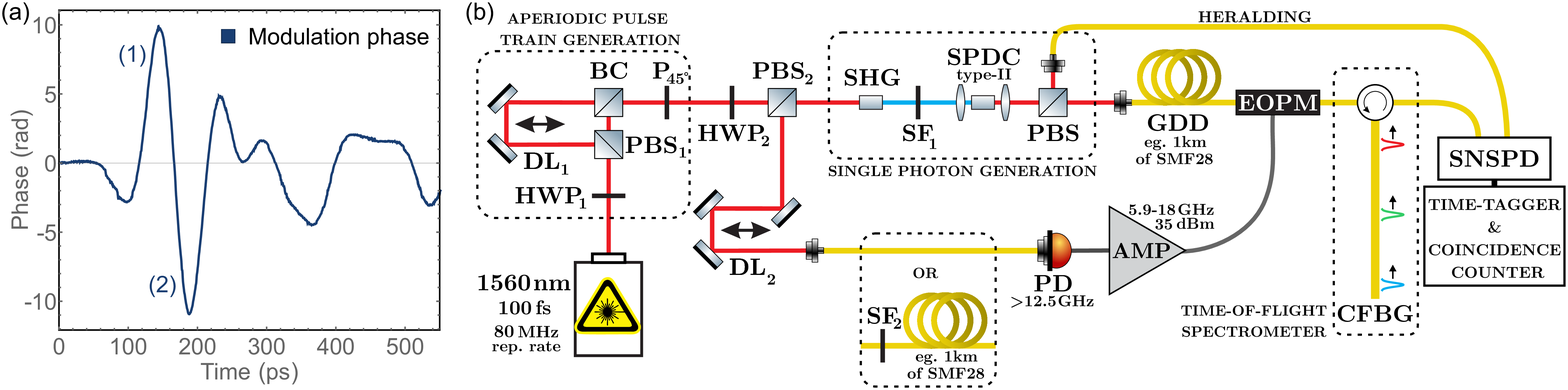}
\caption{\label{fig:schemat} (a) The phase waveform used for time-lensing. One can use either of the two parabolic extrema. Parabola no.\ \textcolor{knof1}{(1)} needs to be combined with normal dispersion, e.g.,\ SMF-28 optical fiber. Parabola no.\ \textcolor{knof1}{(2)} requires anomalous dispersion, e.g.,\ a dispersion compensating fiber. The phase waveform was measured using the electro-optic sampling technique \cite{Wu95,Jachura18}. (b) Experimental setup, see text for details. }
\end{figure*}

Here we use a more direct method for time lens generation, taking advantage form the fact that almost any extremum can be approximated by a parabola. The RF signal driving the EOPM is generated directly from an optical reference by detecting it with a fast photodiode and amplifying, as shown in Fig.~\ref{fig:ogolny}~(b). The extremal parts of the signal are parabolic to a very good approximation, as can be verified in Fig.~\ref{fig:schemat} (a), and are used to realize the time lens. The optical reference is an optical signal derived from the pulse train being modulated, which does not need to fulfill the standard clock signal criteria. Our method is inspired by the previous work on aperiodic electro-optic frequency shifting \cite{Poberezhskiy03}. Combined with an optical or electronic delay line $\Delta t$  this method provides intrinsic synchronization with very low timing jitter and can be used for modulating aperiodic signals, in principle even randomly generated train of pulses, provided an appropriate optical reference signal is available. In such a case, the timing reference has to acquire the same delay as the signal being modulated.

\begin{figure}[b]
\includegraphics[width=\linewidth]{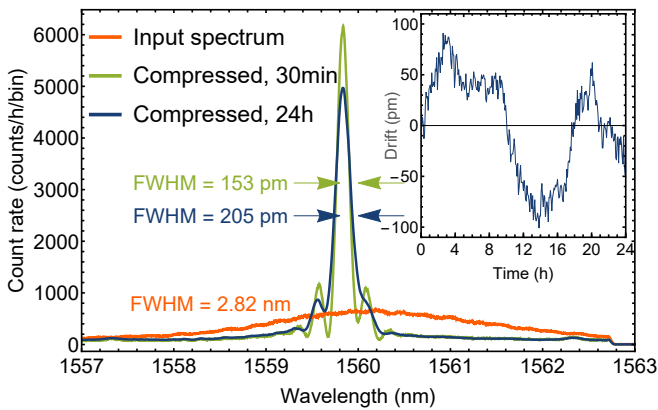}
\caption{\label{fig:kwantowe} Spectra of non-modulated (\textcolor{knof2}{orange}), modulated single-photon pulses acquired in 30 minutes (\textcolor{knofgreen}{green}, $2\times10^{6}$ total coincidence counts) and 24\,hours (\textcolor{knof1}{blue}, $10^{8}$ total coincidence counts), in both cases using a non-delayed optical reference, and without any feedback loop. The spectral bin width is set to 1~pm. The inset shows the drift of the peak of the spectrum measured at 5\,minute intervals. The maximal observed drift rate was 60\,pm/h. It is associated with variation of the environmental conditions in the laboratory.}
\end{figure}

\begin{figure*}
\includegraphics[width=0.99\textwidth]{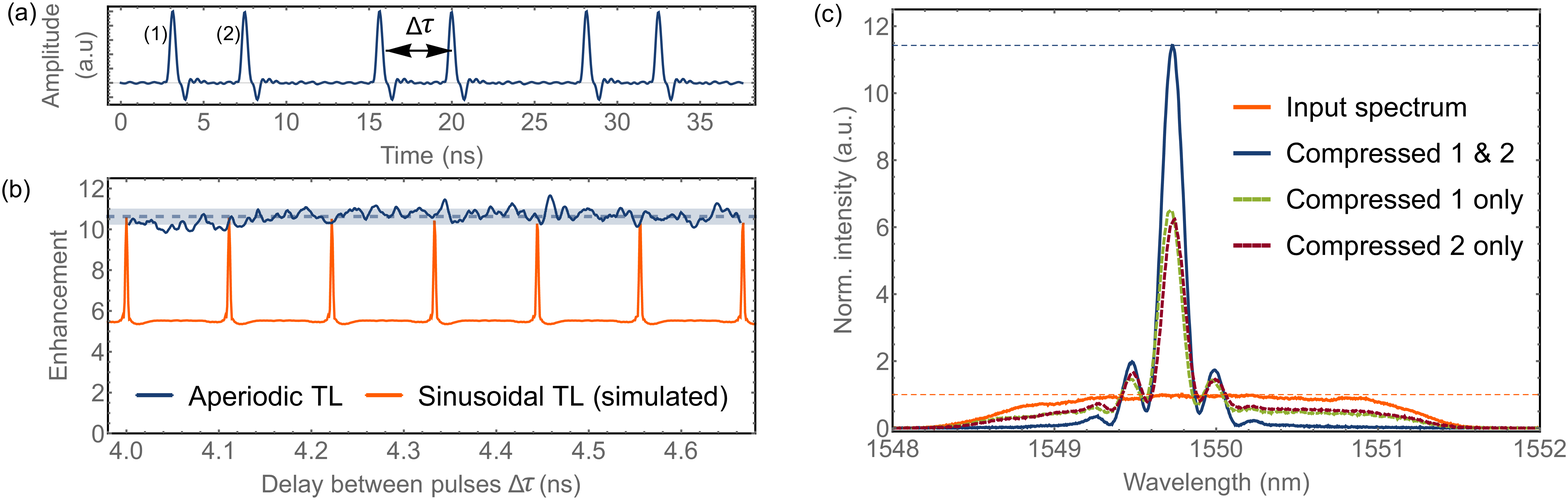}
\caption{\label{fig:aperiodic} (a) Oscilloscope trace of the double pulse train  monitored with a photodiode. Delay $\Delta \tau$ is defined as the delay between two pulses from a single pair. (b) The enhancement (ratio of spectral intensity peaks with and without modulation, see text) obtained by using the aperiodic time lens (experimental values) as compared to the standard sinusoidal time lens with matched parameters, $f_{\color{red}\mathrm{RF}} = 9\;\mathrm{GHz}$ (simulation). The spikes in enhancement for the sinusoidal time lens are achieved for $\Delta \tau$ being multiples of the sine period $f_{\color{red}\mathrm{RF}}^{-1}$. The blue dashed line and blue shade shows the average and standard deviation of enhancement of the aperiodic time lens. (c) The output spectra with modulation turned off (\textcolor{knof2}{orange}), with modulation applied to both pulses (\textcolor{knof1}{blue}), or to only on one of them (\textcolor{knofgreen}{1 -- green}, \textcolor{knofpurple}{2 -- purple}). }
\end{figure*}

In our experiment we use a fast photodiode (PD, EOT ET-3500F, >12.5~GHz bandwidth) as the RF signal source, as shown in the schematic of the experimental setup in Fig.~\ref{fig:schemat}(b). It is preceded by a multipass 10~ns adjustable optical delay line for synchronization (DL$_2$). An additional delay, e.g.,\ a~long optical fiber, may also be introduced, as will be seen in subsequent paragraphs. The duration of the optical pulses detected by the photodiode is much shorter (100~fs) than the inverse of the photodiode electronic bandwidth, such that the generated electronic pulse is fully determined by the photodiode impulse response. Finally, the RF signal is amplified by a high-power RF amplifier (AMP, MiniCircuits ZVE-3W-183+, $5.9$--$18$ GHz bandwidth) to achieve a high level of modulation amplitude necessary to achieve a significant degree of spectral bandwidth compression. Here we set the power of the optical reference such that the amplifier works in saturation. It allows obtaining the highest modulation amplitude possible and largely eliminates the influence of optical power fluctuations of the optical reference.

We will first describe the application of the time lens to single-photon spectral bandwidth compression. The optical pulses are generated with an erbium femtosecond oscillator (Menlo Systems C-Fiber HP, $f_{\mathrm{L}}=80~\mathrm{MHz}$).  For single-photon spectral manipulation, all of the optical power of the laser is guided directly to HWP$_{2}$. Here a polarizing beam splitter (PBS$_2$) is used to separate the optical reference from the signal beam. The optical reference is used to generate the time lens signal, as described above.

The remaining signal beam is then frequency-doubled in a second harmonic generation module (SHG, Menlo Systems), spectrally filtered (SF$_{1}$), and used to drive type-II collinear spontaneous parametric down-conversion in a bulk 10-mm-long periodically poled potassium titanyl phospahate (PPKTP) crystal. Then the generated pair of orthogonally polarized signal and idler photons is split with PBS$_{3}$ and fiber coupled. The idler is directed straight to a single-photon detector for heralding. We spectrally filter the pump to achieve uncorrelated joint spectral amplitude of the photon pair, which is necessary for heralding single photons in a spectrally pure state \cite{Shi08,Mosley08}.  The signal photon undergoes group delay dispersion (GDD) in $1$~km of standard SMF-28 optical fiber, which is followed by the electro-optic phase modulator (EOPM, EO-Space, $16$~GHz bandwidth, $V_{\pi}=3\,\mathrm{V}$\,@\,$1$~GHz), which acts as the time lens. 

The temporal phase waveform applied to the signal pulses passing through the EOPM, shown in Fig.~\ref{fig:schemat}(a), was measured by direct electro-optic sampling \cite{Wu95,Jachura18} using bright classical light. It allowed measuring the introduced phase shot-by-shot, which enabled obtaining information about the timing jitter. We conservatively estimate the timing jitter between the RF and optical signals to be below 30~fs. Such low timing noise is characteristic of microwave photonic approaches \cite{Capmany07,  Quinlan13}.

To realize the time lens we synchronize the single-photon pulses with parabola no.\ (1), shown in Fig.~\ref{fig:schemat}(a), and use 1~km of SMF-28 to introduce GDD of 22.4~ps$^{2}$. To reach the truly aperiodic operating mode, where the pulse used to generate a single photon and the pulse used to generate the RF signal modulating the single photon are derived from the same laser pulse, a matching delay of the optical reference may be introduced in the form of $1$~km of SMF-28 fiber. Then the bandwidth of the optical reference needs to be limited to $1.5$~nm, using the adjustable spectral filter SF$_2$, to prevent modification of the photodiode RF output due to dispersive broadening of the optical reference pulse. However in our setup, where single-photon pulses are produced at multiples of $1/f_{\mathrm{L}} = 12.5$~ns intervals, the $1$~km delay is not essential.

To measure the spectra of signal-photon pulses, we employ dispersive single-photon spectrometry \cite{Davis17}. We use a chirped fiber Bragg grating module (CFBG, Proximion) with dispersion of $5$~ns/nm, which maps the spectrum to the time-of-arrival of single photons. Both signal and idler photons are then detected with niobium nitride superconducting nanowire single-photon detectors (SNSPD, Single Quantum, $50\%$ system detection efficiency, $20$~ps full-width-half-maximum timing jitter) and time-tagged for time-resolved coincidence counting using Picoquant HydraHarp 400 time tagger. This yields the heralded spectrum of the signal photon wave packets. The resolution of the spectral measurement is limited by the CFBG dispersion and is better than $45$~pm \cite{Goda09}.

First, we show bandwidth compression of single photons using a periodic pulse train (one pulse per repetition period) and non-delayed detection of the optical reference for RF generation. The truly aperiodic operating mode was not used due to limited optical power available in our setup.  The joint spectral intensity of photon pairs was set to symmetrical (uncorrelated) to achieve pure heralded single photons. In Fig.~\ref{fig:kwantowe}, we plot the acquired spectra of the signal photon before turning on the time lens (orange) and with the time lens turned on and synchronized with parabola (1), acquired during an integration time of $30$~minutes (green) and $24$~hours (blue), respectively. The achieved compressed spectral width for shorter measurement was $153$~pm with an enhancement of $9$, where the enhancement is defined as the ratio between the maximal intensities of the compressed and input spectra. For the long measurement, it was $205$~pm width and enhancement of $7$. The broadening of the spectrum is related to the long-term drifting of the maximum, which arises from environment instability. Here, thanks to isolating dispersion fibers and water-cooling the EOPM, the remaining drift is very slow, with a maximum rate of $60$~pm/h. It is associated with slow variation of the environmental conditions, most likely via inducing changes of the laser repetition rate, resulting in drift due to the use of non-delayed detection of the optical reference in the single-photon experiment. The drift within the $24$ hours is shown in the inset of Fig.~\ref{fig:kwantowe}, with a timing interval of $5$ minutes. This is a significant stability improvement as compared to initial experiments on electro-optic single-photon spectral modifications, where the stability was on the order of single minutes \cite{Karpinski17}. The stability could be further improved by locking the repetition rate of the laser to an external reference or by using the truly aperiodic scheme with a delayed optical reference.

We experimentally  verified  the  correct  operation  of  the  time  lens in the  truly  aperiodic  configuration  by performing spectral bandwidth compression with bright classical light (bypassing single-photon generation), while varying the repetition rate of the laser by $\pm200$~Hz. We observed a stable spectrum at the output, whose central wavelength nor position did not vary within the 20 pm resolution of the optical spectrum analyzer used for detection (OSA, not shown in Fig. 2). This indicates that, in a general context, the presented aperiodic time lens could be used to modulate single-photon pulses in schemes with arbitrary timing, such as in quantum networks. 

The maximum pulse duration at the time lens is limited by the width of the central peak of the phase waveform shown in Fig.\ \ref{fig:schemat}(a). The parts of pulses that fall outside this temporal aperture are subject to aberrations \cite{Kolner94}, whose traces can be seen in the features surrounding the central peaks in Fig.\  \ref{fig:kwantowe}. The time lens could be made to accept longer pulses by broadening the central peak of the phase waveform, e.g. by low-pass filtering the PD output. However this would not change the achievable enhancement, because the spectral widths of the input and output would need to be scaled accordingly. The enhancement is primarily determined by the phase modulation amplitude\cite{Karpinski17}, which is limited in our case by the saturation of the RF amplifier, and in general by the breakdown voltage of the EOPM\cite{Torres11}. The enhancement could be increased by broadening the temporal aperture through the use of more complex phase modulation patterns as discussed in ref. \cite{Sosnicki18}.

We demonstrate aperiodic operation of the time lens using classical coherent light. To generate double pulses, we split the laser output beam with a polarizing beamsplitter (PBS$_{1}$) into short and long beam paths, where the latter includes a variable delay line (DL$_{1}$).  The beams are then recombined with a beam combiner (BC) and projected into the same polarization mode with a linear polarizer at 45$^{\circ}$ (P$_{45^{\circ}}$). Setting a half-wave plate (HWP$_{1}$) before the polarizer allows choosing between the generation of single or double pulses with a variable delay. Then using a half-wave plate (HWP$_{2}$) and PBS$_{2}$ part of the optical power is picked up for the optical reference. Single-photon generation is bypassed, and spectra are measured using the OSA (not shown in Fig.~2).

We set HWP$_{1}$ to produce double pulses with an adjustable delay $\Delta\tau$ between them. An oscilloscope trace of the pulse train measured with a $1$~GHz photodiode is shown Fig.~\ref{fig:aperiodic}(a). For this part of the experiment we used also the aforementioned additional delay of the optical reference. Then across the delay range $\Delta \tau$ of $0.66$~ns, we measured the enhancement of bandwidth compression, with the results shown in Fig.~\ref{fig:aperiodic}~(b). One can see a nearly constant enhancement over the whole range of delays, which proves the aperiodicity of our time lens. In contrast, we also show a simulated enhancement for a sinusoidal time lens with parameters matching the parabola no. (1) of the modulating pulse we use, which are $f_{\mathrm{RF}}=9\;\mathrm{GHz}$ and $A = 10\;\mathrm{rad}$. The sinusoid is then synchronized with the first pulse from the pair, hence the average enhancement here is half of the enhancement for the aperiodic time lens: in these regions, only one of the pulses acquires a proper quadratic phase. The observed spikes in the enhancement are separated by 0.11~ns and observed where the delay between the pulses matches the multiples of the period of RF signal $\Delta \tau = n f_{\mathrm{RF}}^{-1}$, where $n$ is an integer. In Fig.~\ref{fig:aperiodic}~(c) we show measured spectra of input pulses (orange) and spectrally compressed pulses when both pulses are modulated (blue), and when only one of the pulses is modulated, the first or the second one respectively (green, purple).

We experimentally verified the minimum delay between two pulses that can be time-lensed by finding the value of $\Delta\tau$, for which the enhancement was reduced by more than 10\% with respect to the level indicated in Fig.~\ref{fig:aperiodic}(b), to be 800~ps. It is limited by the tail of the applied phase waveform and could be reduced by increasing the RF amplifier bandwidth.

In summary, we have experimentally demonstrated an electro-optic time lens capable of modulating aperiodic pulse trains, whenever there is a synchronized timing reference available. By driving the electro-optic phase modulator with a signal obtained by direct detection of the optical reference, a very low timing jitter value of <30~fs  was obtained. We show a stable compression of heralded single photons with enhancement exceeding 9 over 30~mins, and exceeding 7 over 24~hours, obtained with no feedback of any kind. We explicitly show the aperiodic operation of the proposed time lens by compressing the spectral bandwidth of a double-pulse train with a variable delay between the pulses. In principle, the demonstrated time lens may be used for modulating a random train of optical pulses. We anticipate that this method, due to its simplicity and stability, will extend the application of the electro-optic approach to spectral-temporal shaping of single-photon pulses for quantum information processing, especially in the context of spectral-temporal mode matching for quantum networks. Further we expect this approach to enable building electro-optic spectro-temporal optical systems, where multiple time lenses and dispersive elements are combined to achieve complex time-frequency transformations of quantum light.

\vspace{-1em}
\begin{acknowledgments}
We would like to thank C. Radzewicz and M. Jachura for insightful discussions and K. Banaszek for access to experimental equipment. This research was funded in part by the National Science Centre of Poland (project no. 2014/15/D/ST2/02385) and in part by the First TEAM programme of the Foundation for Polish Science (project no.\ POIR.04.04.00-00-5E00/18), co-financed by the European Union under the European Regional Development Fund.
\end{acknowledgments}
Experimental data can be obtained by email on request to the corresponding author.

\end{document}